\documentstyle{aa}
\input psfig.sty  
\begin{document}

\date{Version of 18 November 1999} 

\thesaurus{03(
	      11.04.1; 
	      11.04.2; 
	      11.06.2; 
	      11.07.1; 
	      }  

\title{HI observations of nearby galaxies \\ 
       I. The first list of the Karachentsev catalog
}


\author{ W.K.Huchtmeier \inst{1} \and I.D. Karachentsev \inst{2} 
	 \and V.E. Karachentseva \inst{3} \and M. Ehle \inst{4}} 

\offprints{W.K. Huchtmeier; \it email: huchtmeier@mpifr-mpg.de} 


\institute{Max-Planck-Institut f\"{u}r Radioastronomie, Auf dem H\"{u}gel 69,
		    D-53121 Bonn, Germany 
\and Special Astrophysical Observatory, Russian Academy of Sciences,
		    N.Arkhyz, KChR, 357147, Russia
\and Astronomical Observatory of Kiev University, Kiev, Ukraine  
\and Max-Planck-Institut f\"{u}r Extraterrestrische Physik,
Giessenbachstra\ss{}e, D-85740 Garching bei M\"{u}nchen, Germany}

\maketitle

\begin{abstract}
We present HI observations of the galaxies in the first list of the
Karachentsev catalog of previously unknown nearby dwarf galaxies 
(Karachentseva \& Karachentsev 1998). This survey covers all known nearby galaxy groups
within the Local Volume (i.e. within 10 Mpc) and their environment,  
that is about 25\% of the total sky. 
A total of 257 galaxies have been observed with a detection rate of
60\%. We searched a frequency band corresponding to heliocentric radial
velocities from -470 km\,s$^{-1}$ to $\sim$+4000 km\,s$^{-1}$. 
Non-detections are either due to limited coverage in radial
velocity, confusion with Local HI (mainly in the velocity range -140
km\,s$^{-1}$ to +20 km\,s$^{-1}$), or lack of sensitivity for very weak emission. 
25\% of the detected galaxies  are located within the Local Volume.
Those galaxies 
are dwarf galaxies judged by their optical linear diameter (1.4 $\pm$
0.2 kpc on the average), their mean total HI mass (4.6 10$^{7}$
M$_{\odot}$), and
their observed linewidths (39 km\,s$^{-1}$ ).
\end{abstract}

\section{Introduction}
The only way to study the smallest galaxies is to search for them in our cosmic
neighborhood. The first systematic catalog of nearby galaxies was 
prepared by Kraan-Korteweg \& Tammann (1979) who collected all known
galaxies with corrected radial velocities v$_{0}$ $\le$500 km\,s$^{-1}$,
a total of 179 objects (hereafter called the KKT sample). 
Since that time the number of known galaxies
within the Local Volume (i.e. within a distance of 10 Mpc) increased to
303 objects (Karachentsev et al. 1999). 
For the past decade the initial KKT sample has been
increased almost two times in number due to the mass redshift surveys  
of galaxies from the known catalogues, revealing  new nearby galaxies in
the Milky Way "Zone of Avoidance", as well as special searches for dwarf
galaxies in nearby groups. The increasing numbers of galaxies in the Local
Volume is 
mainly due to many new dwarf galaxies. 
This fact demonstrates how incomplete our knowledge about the galaxy 
population of even the Local
Volume is.

A couple of years ago Karachentseva \& Karachentsev (1998; hereafter KK98)
initiated an all-sky search for candidates for new nearby dwarf galaxies
using the second Palomar Sky Survey and the ESO/SERC plates of the
southern sky. The results of the first two segments of the survey have
been published, they cover large areas around the known galaxy groups in the 
Local
Volume (KK98) and the area of the Local Void (Karachentseva et al.
1999).
In a next step to derive distances we will measure radial
velocities. Later on we will aim for more exact photometric distances. 
In this paper we present the first follow-up observations, the HI search
for the galaxies in KK98. The HI search for dwarf irregular galaxies seems
quite efficient as these galaxies are HI rich in general and with
adequate velocity resolution, say 5 km\,s$^{-1}$, all the HI of a given
galaxy will be within a few velocity channels. The characteristic
signature of a dwarf galaxy profile, a nearly gaussian structure, 
is different from radio interference and
easily will lead to a good signal-to-noise ratio.

\section{Observations}
Observations were performed with three different radio telescopes 
for different declination ranges. The 100-m radiotelescope at Effelsberg
was used for declinations greater than  $-31\degr$, the Nan\c{c}ay radio
telescope was selected for galaxies in the declination range $-38\degr \le
-31\degr$, and the compact array of the Australia Telescope was used for
galaxies south of $-38\degr$.

\subsection{Effelsberg observations} 
The radio telescope at Effelsberg has been used in the total power mode
(ON -- OFF) combining a reference field 5 minutes earlier in R.A. with the
on-source position. A dual channel HEMT receiver had a system noise of
30K. 
\newpage  
The 1024 channel autocorrelator was split into 4 bands (bandwidth 6.25 MHz)
of 256 channels each
shifted in frequency by 5 MHz with respect to their neighbor in order to 
cover a velocity range from -470 to 3970 km\,s$^{-1}$ overlapping 1.5
MHz between channels.
The resulting channel separation was 5.1 km\,s$^{-1}$ yielding a
resolution of 6.2 km\,s$^{-1}$ (10.2 km\,s$^{-1}$ after Hanning
smoothing). The HI profiles observed with the 100-m radiotelescope 
are presented in Fig.\,1 in order of increasing R.A. as in Table\,1. 
The half power beam widths (HPBW) of the Effelsberg
telescope at this wavelength is 9\farcm3.

\subsection{Nan\c{c}ay observations} 
For 15 galaxies in the declination range $-38\degr \le -31\degr$ 
the Nan\c{c}ay radio telescope was used with the same velocity
resolution and coverage. Major differences to the description given for
the Effelsberg observations were a different system noise (45K), a 
different antenna beam ($3\farcm 6 \times 22 \arcmin$ in R.A. and Dec. 
for this declination range), and shorter 
integration phases with a cycle of 2 minutes for the ON and the OFF 
positions. Nine galaxies have  been detected (Fig.\,2).

\begin{figure*}[ht] 
\psfig{figure=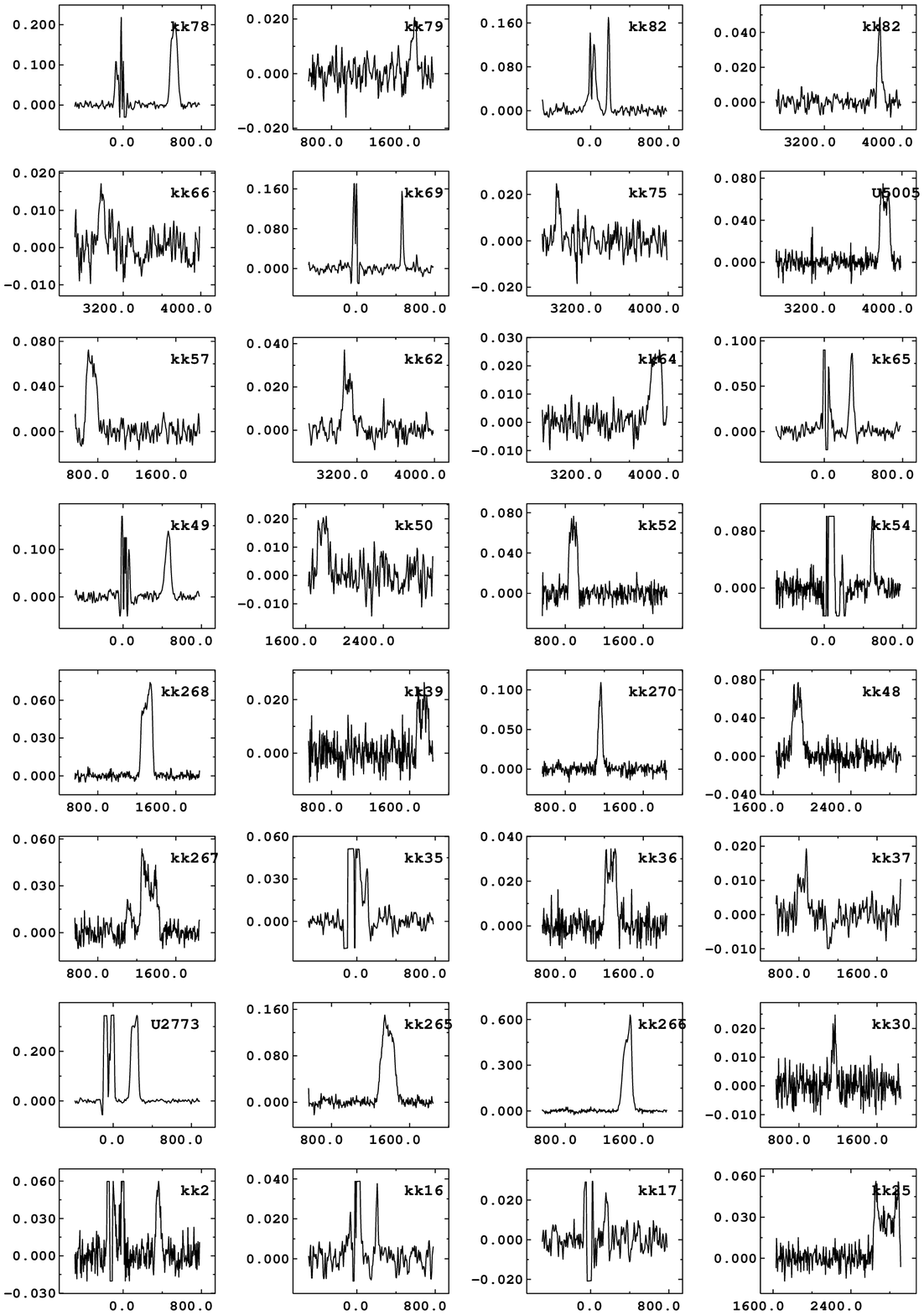,width=15cm} 
\caption{
   HI profiles observed with the 100-m radio telescope at Effelsberg 
   which has a HPBW of 9\farcm3 at a wavelength of 21~cm.
   Observations were obtained in the total power mode [ON -- OFF] which
   yields a residual of the Local HI emission around 0 km\,s$^{-1}$.   
   The profiles are arranged in ascending R.A. starting at the bottom
   left corner.}
\addtocounter{figure}{-1}
\end{figure*} 
\begin{figure*}[ht] 
\psfig{figure=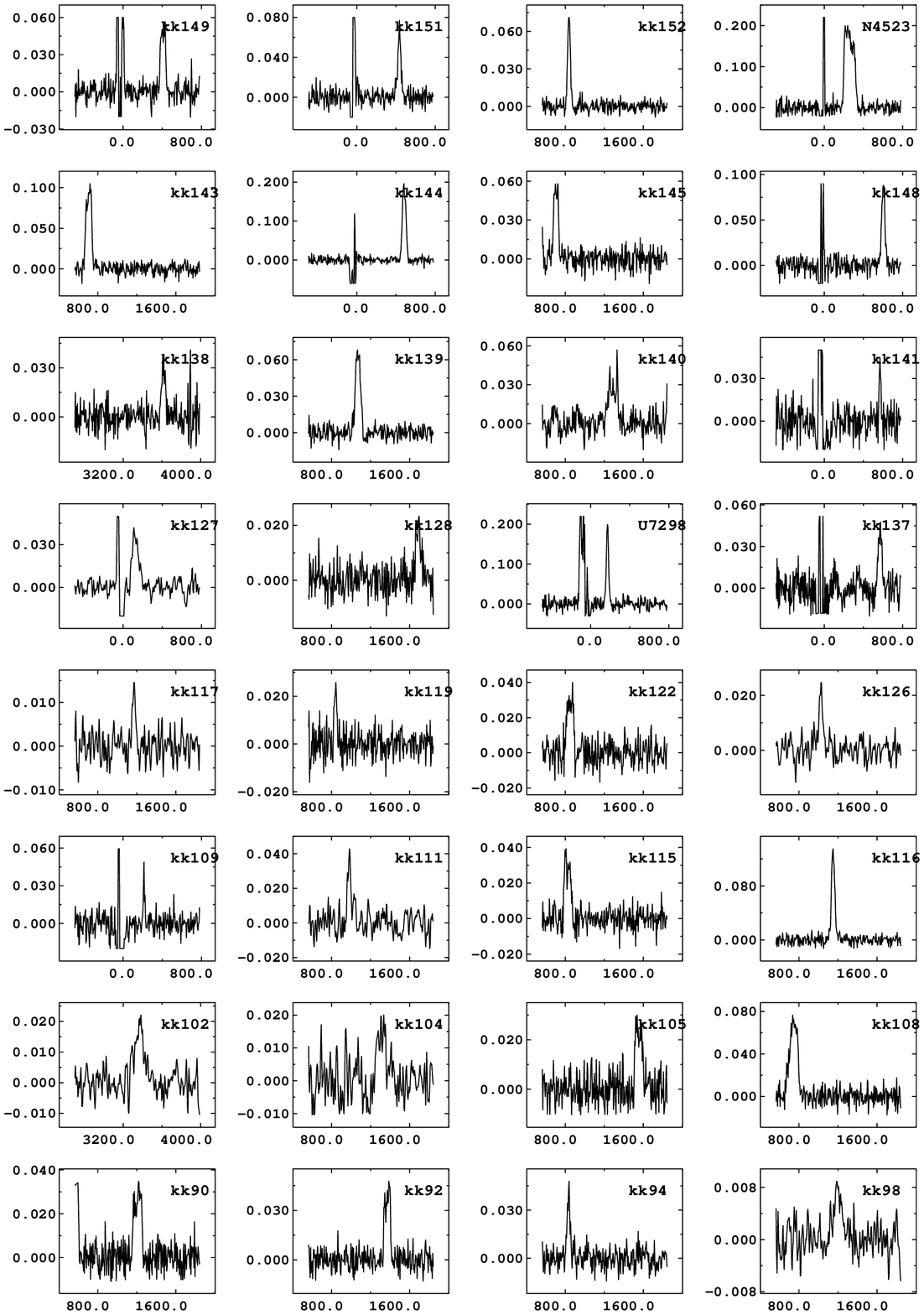,width=15cm}
\caption{cntd.}
\addtocounter{figure}{-1}  
\end{figure*} 
\begin{figure*}[ht] 
\psfig{figure=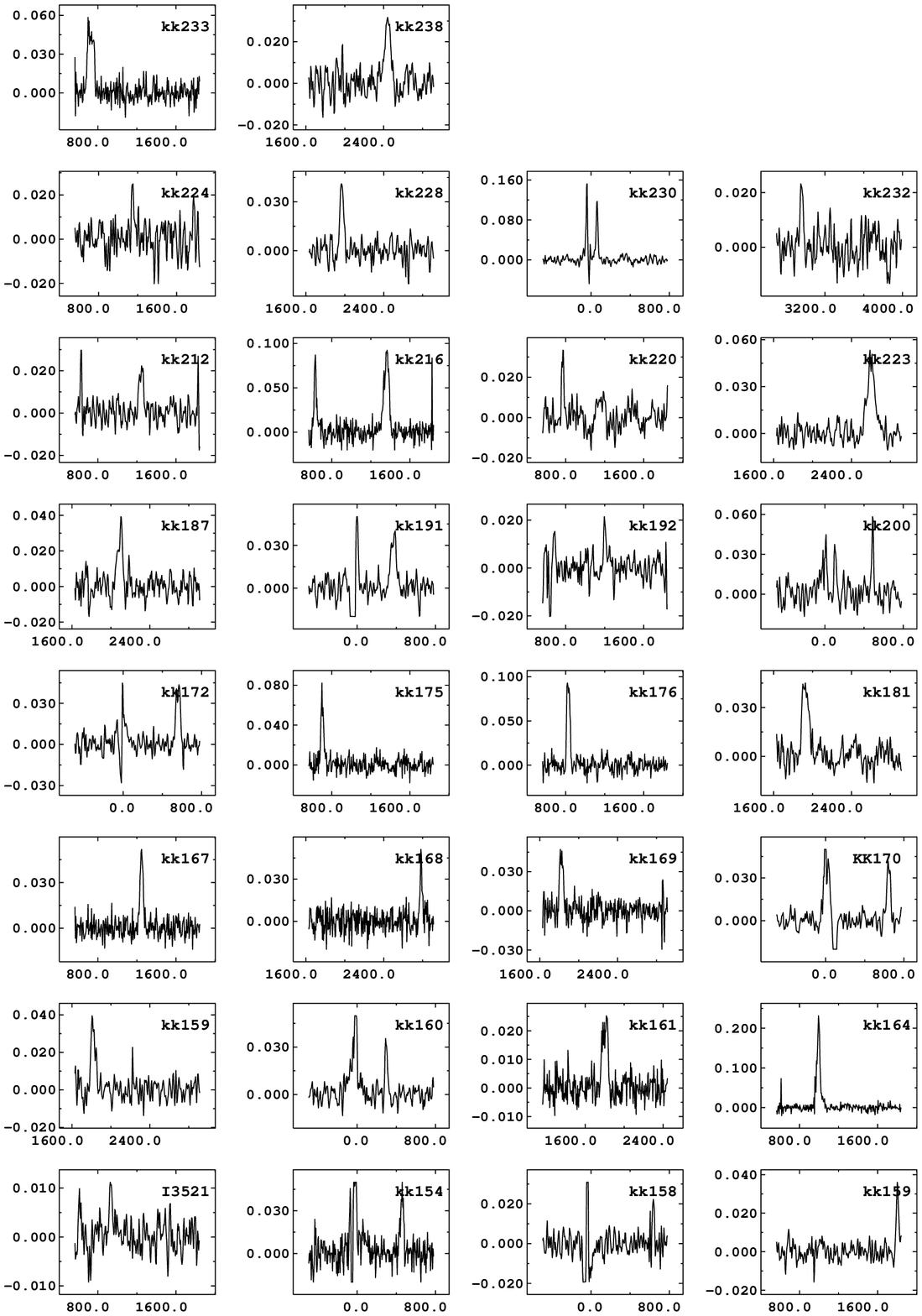,width=15cm}
\caption{cntd.}
\end{figure*} 

\subsection{Compact Array of the Australia Telescope}  
40 of the 57 galaxies south of declination $-38\degr$ have been observed
with the Compact Array of the Australia Telescope.
For this HI search we have chosen the 750A antenna array configuration 
in order 
to yield an antenna beam comparable to the optical size
of the smallest galaxies (i.e. $\sim$ 1$\arcmin$). The frequency setup and
correlator configuration was such that we obtained a velocity coverage 
from -450 
to +2900 km\,s$^{-1}$ and a channel separation of 6.6 km\,s$^{-1}$ 
(i.e. a resolution of 7.9 km\,s$^{-1}$). 
Each galaxy was observed for 10 min every few hours. With five to six
observations per target position we achieved a regular coverage of the 
uv plane for these
'snapshot mode' observations.
The resulting integrated HI profiles are given in Fig.\,3 (for a more
detailed discussion of these data see Huchtmeier et al. in preparation).
We may miss some flux with the interferometer (missing flux) as the
observed HI emission extends over more than 2$\arcmin$ per channel for
over 60\% of the galaxies. 
Galaxies from the kk98 sample not observed are: kk\,11, kk\,63, kk179, 
kk\,184, kk\,189,
kk\,190,
kk\,197, kk\,203, kk\,211, kk\,213, kk\,214, kk\,217, kk\,221, kk\,222,
kk\,235, kk\,244, kk\,248.  

\begin{figure*}[ht] 
\psfig{figure=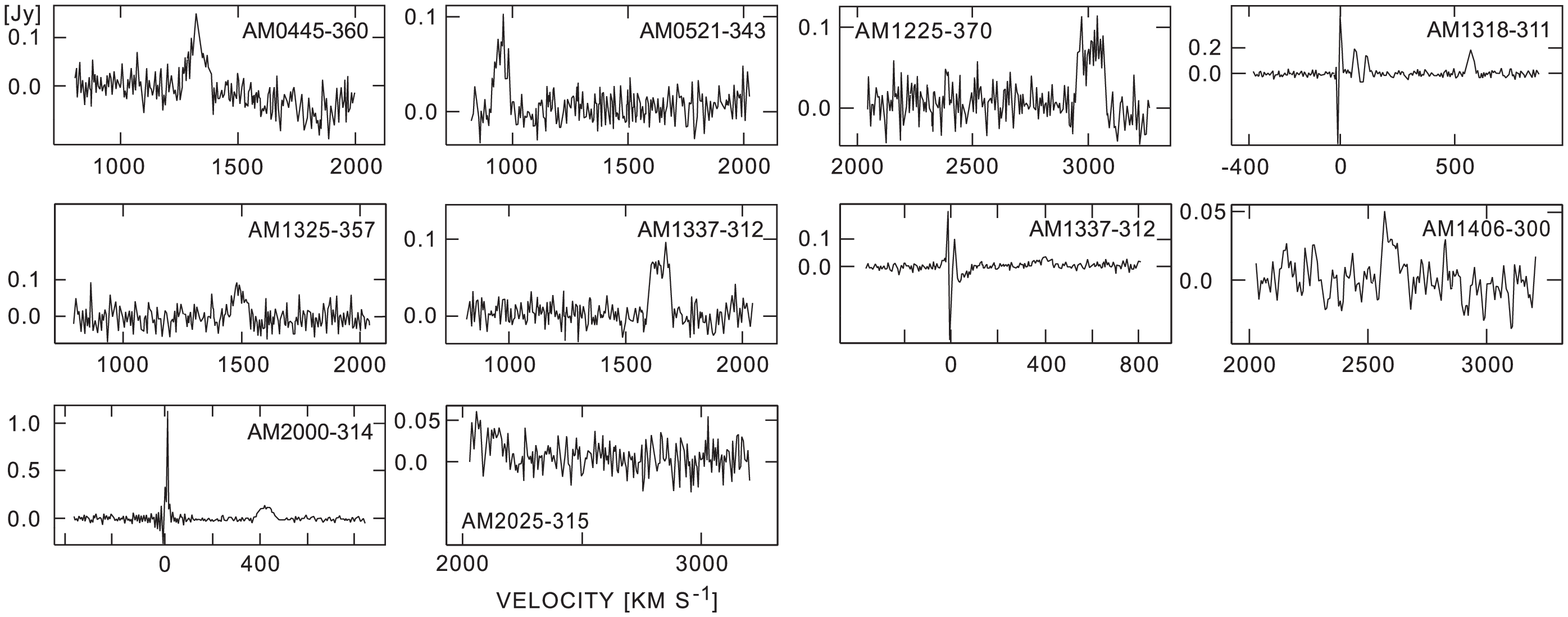,width=12cm} 
\caption{
     HI profiles observed with the Nan\c{c}ay radio telescope (HPBW of 
     $3\farcm6 \times 22\arcmin$ for the declination range in question)
       }
\end{figure*} 

\begin{figure*}[ht] 
\psfig{figure=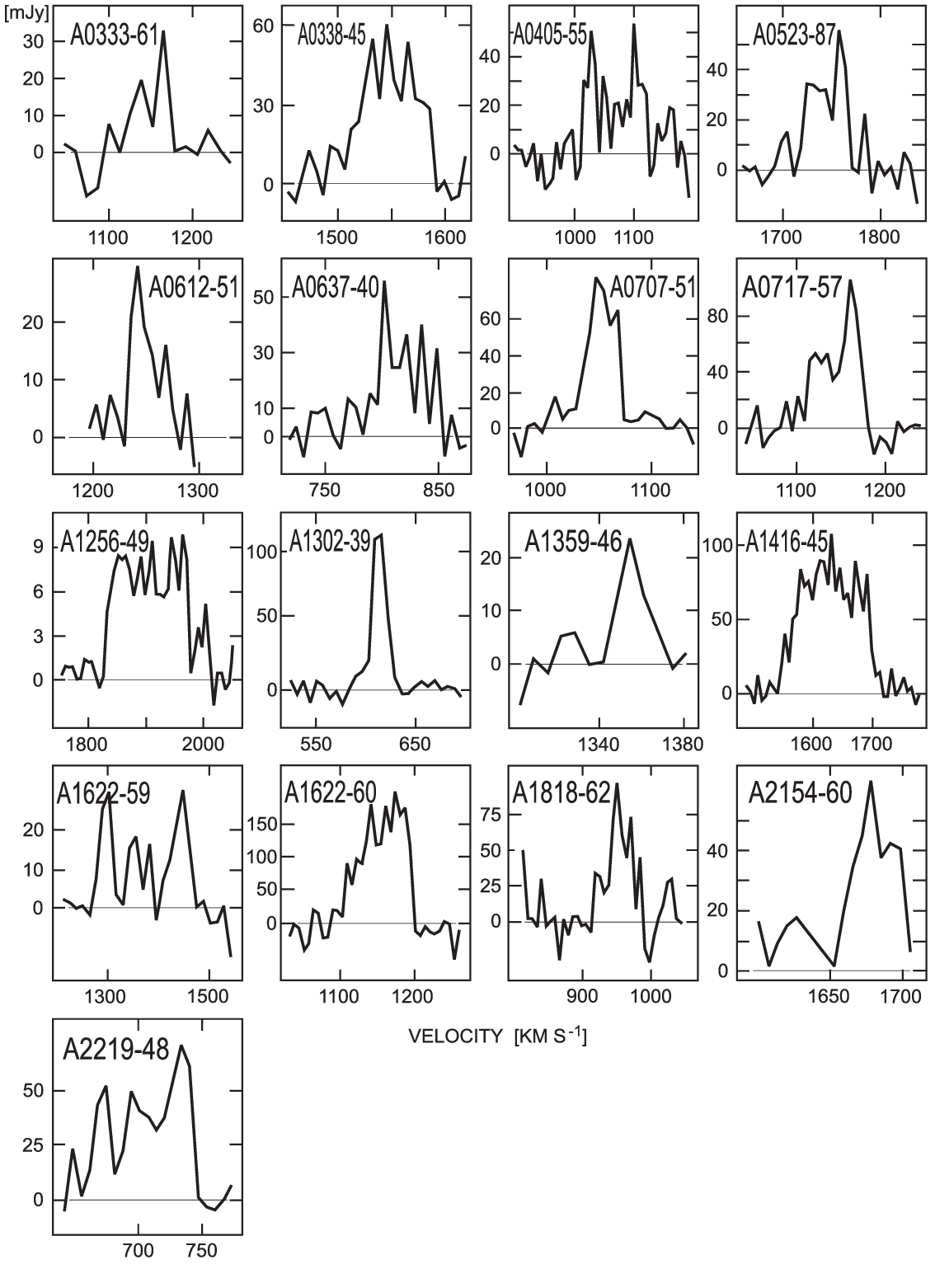,width=12cm} 
\caption{
      HI profiles observed with the Australia
      Telescope Compact Array. The synthesized antenna beam is of the order 
      of 1$\arcmin$
       }
\end{figure*} 

\section{The data}
Our search list was an early version of the list of KK98 containing 
a few additional galaxies which did not make it into the final version 
because of their morphology and/or size (i.e. they were too small).
Particularly, we took into account the results of HI searches for
nearby dwarf galaxies made by Kraan-Korteweg et al. (1994), Huchtmeier
et al. (1995), Burton et al. (1996), Huchtmeier \& van Driel (1997),
Huchtmeier et al.(1997) and Cote et al.(1997).
The optical data of our galaxies are given in Table\,1.
The kk-number (or other identification if there is no kk-number) is
given in column 1, R.A. and Dec. (1950) follow in columns 2 and 3. 
The optical
diameters $a$ and $b$ in the de Vaucouleurs ($D_{25}$) system follow in
columns 4 and 5, the morphological type in column 6 where we use the following
coding:
\par Im - irregular blue object with bright knot(s),
\par Ir - irregular without knots or with amorphous condensations, the
colour is neutral or bluish,
\par Sm - disturbed spiral or irregular with signs of spiral structure,
\par Sph - spheroidal, with very low brightness gradient or without any,
the color is neutral or redish.

The optical surface brightness (SB) has been coded (see KK98):
high (H), low (L), very low (VL), and extremely low (EL) in column 7.
The total blue magnitude B$_{t}$ and its reference follow in columns 8
and 9. 
'NED' - data are from the NASA/ Extragalactic Database, 'IK' -  visual
estimates from POSS (typical error is about 0.4 mag) by I.
Karachentsev, '6m' - accurate photometric data from the 6-m telescope
CCD-frames obtained by Karachentsev and coworkers (unpublished); 
'UH' - photometric data from U. Hopp (Calar Alto) unpublished. 
The Galactic extinction follows in column 10. Other names
(identifications) are listed in column 11.

\begin{figure}[ht] 
\psfig{figure=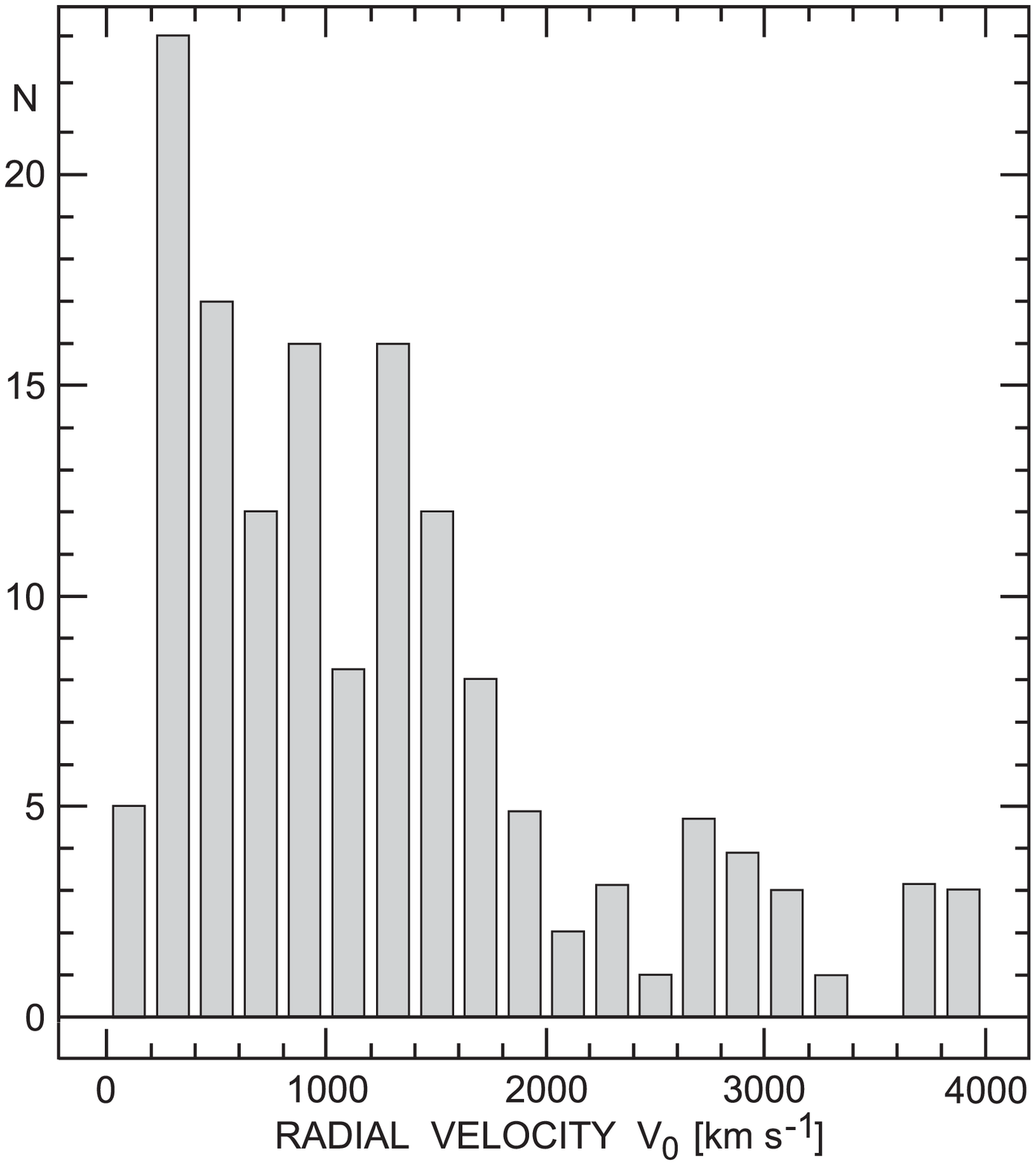,width=8.8cm} 
\caption{
      The histogram shows the number of galaxies per velocity interval
      of 200 km\,s$^{-1}$. The distribution of corrected radial
      velocities ($v_{0}$) of our galaxy sample demonstrates the local
      character of these galaxies
       }
\end{figure}  

Results of the HI observations are summarized in Table\,2.
The kk-number is given in column 1, the HI-flux [Jy\,km\,s$^{-1}$]
follows in column 2, the maximum emission and/or the r.m.s. noise  [mJy] 
in column 3, the heliocentric radial velocity plus error in column 4, 
the line widths
at the 50\%, the 25\%, and the 20\% level of the peak emission in column
5. Distances (column 6) have been derived with different methods, there are 
photometric distances in some cases, in other cases the group membership
yields a distance. If no other distance estimate is available, we
assumed a Hubble constant of 75 km\,s$^{-1}$Mpc$^{-1}$ to derive a 
'kinematic' distance. The absolute magnitude is given in
column 7, the integrated HI mass (column 8)  
was calculated as (e.g. Roberts 1969)  

 $$ (M_{HI}/M_{\sun})= 2.355 \times 10^{5} \times D^{2} \times \int
 S_{v}dv $$

where $D$ is the distance of the galaxy in Mpc and $\int S_{v}dv$ is the
integrated HI flux in Jy\,km\,s$^{-1}$. 
The relative HI content $M_{HI}/L_{B}$ follows in column 9. 
Finally, column 10 contains comments relative to the telescope used
for the observation: unless otherwise noted observations have been
performed with the 100-m radiotelescope at Effelsberg, N - marks the
Nan\c{c}ay radio telescope, ATCA - the Australia Telescope Compact Array
at Culgoora, NSW.
\rm 
In a number of cases emission at negative radial velocities has been
observed (kk20, kk236, kk237; only kk\,236 has been plotted as an
example). The Dwingeloo HI survey (Hartmann \& Burton 1997)
has been consulted: in all cases of negative radial velocities 
extended HI emission was found suggesting that we observed high velocity 
clouds in our Galaxy.

\begin{figure}[ht]
\psfig{figure=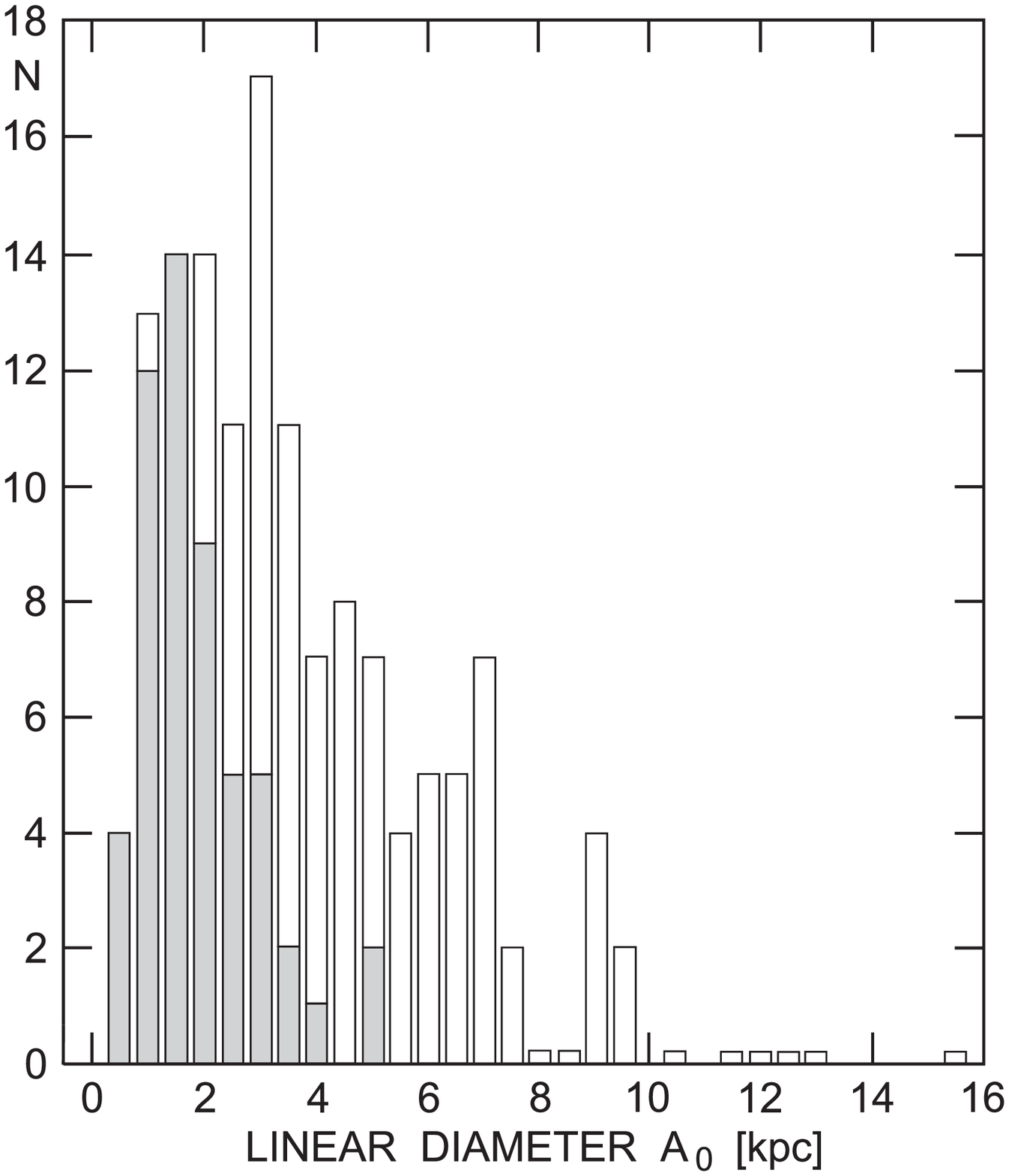,width=8.8cm}  
\caption{
     The distribution of the optical linear diameter $A_{0}$ in kpc for the
     whole sample in the de Vaucouleurs ($D_{25}$) system is given here.
     Galaxies within 10 Mpc (i.e. within the Local Volume) are shown
     by shaded areas. The medium value for the shaded areas is 
     $1.4 \pm 0.2$~kpc
       }
\end{figure}

\section{Discussion} 
A great majority (73\%) of our galaxies are of type Im (26) and Ir
(162), about 20\% are of type Sph/Ir (12) and Sph (39), while the 
rest of 8\% is a collection of different types from spiral to Im/Sm and BCD. 
The detection rate of our sample galaxies depends on the morphological type. 
75\% of the spirals (type S0 to Sm/Im and BCD) were detected; 
the detection rate for types Im and Ir is very similar close to 60\%,
whereas the detection rate for types Sph/Ir and Sph is considerable lower at 33
and 23\%, respectively.
The detection rate depends on the optical surface brightness (SB) class, too.
From high SB to low, very low, and extremely low SB the detection rate 
decreases from 70\% to 58\%, 49\%, and 43\%, respectively. This trend
reflects the type dependence and the fact that we deal with fainter
galaxies as we descend from high SB to very low SB, the median absolute
magnitudes for the detected galaxies change from -15.43 (H) to -13.92
(VL) for our brightness classes. 
 
A number of the galaxies within the present sample are associated with
nearby groups of galaxies (e.g. Tully 1988) according to their position, radial
velocity and relative resolution: 
   
NGC\,672 group: kk\,13, kk\,14, kk\,15; 

NGC\,784 group: kk\,16, kk\,17; 

Maffei group: kk\,19, kk\,21, kk\,22, kk\,23, kk35\, kk\,44;

Orion group: kk\,49; 

M\,81 group: kk\,81, kk\,83, kk\,85, kk\,89, kk\,89, kk\,91; 

Leo group: kk\,94; 

CVn cloud: kk\,109, UGC\,7298, kk\,137, kk\,141, kk\,144, kk\,148, 
kk\,149, kk\,151, kk\,154, kk\,158, kk\,160, kk\,191, kk\,206,
kk\,220, kk\,230; 

Centaurus group: kk\,170, kk\,179, kk\,182, kk\,190, kk\,191,
kk\,195, kk\,197, kk\,200, kk\,211, kk\,217, kk\,218; 

NGC\,6946 group: kk\,250, kk\,251, kk\,252; 

Virgo cluster: kk\,111, kk\,127, kk\,128, kk\,140, NGC\,4523,
IC3517, kk\,164, kk\,168, kk\,169, kk\,172, kk\,173, U\,8091.

There are a few cases of high $M_{HI}/L_{B}$ values in Table\,2.
Four of the five galaxies with $M_{HI}/L_{B} \ge 5$ are actually found to be
confused by emission from nearby galaxies (see footnotes to Table\,2). 

The present sample of galaxies as presented in Tables 1 and 2 will be
discussed now in some detail with the help of global parameters.
The distribution of radial velocity (v$_{0}$, corrected for the rotation
of our galaxy)  is given in Fig.\,4. Apart from a few
background objects most of the galaxies belong to the local
supercluster, about 25\% are within the Local Volume.
 From this situation it is clear that the great majority of the galaxies in
the present sample are dwarfish in nature. This will be shown more
convincingly below when we compare several other global parameters of these 
objects.

\begin{figure}[ht] 
\psfig{figure=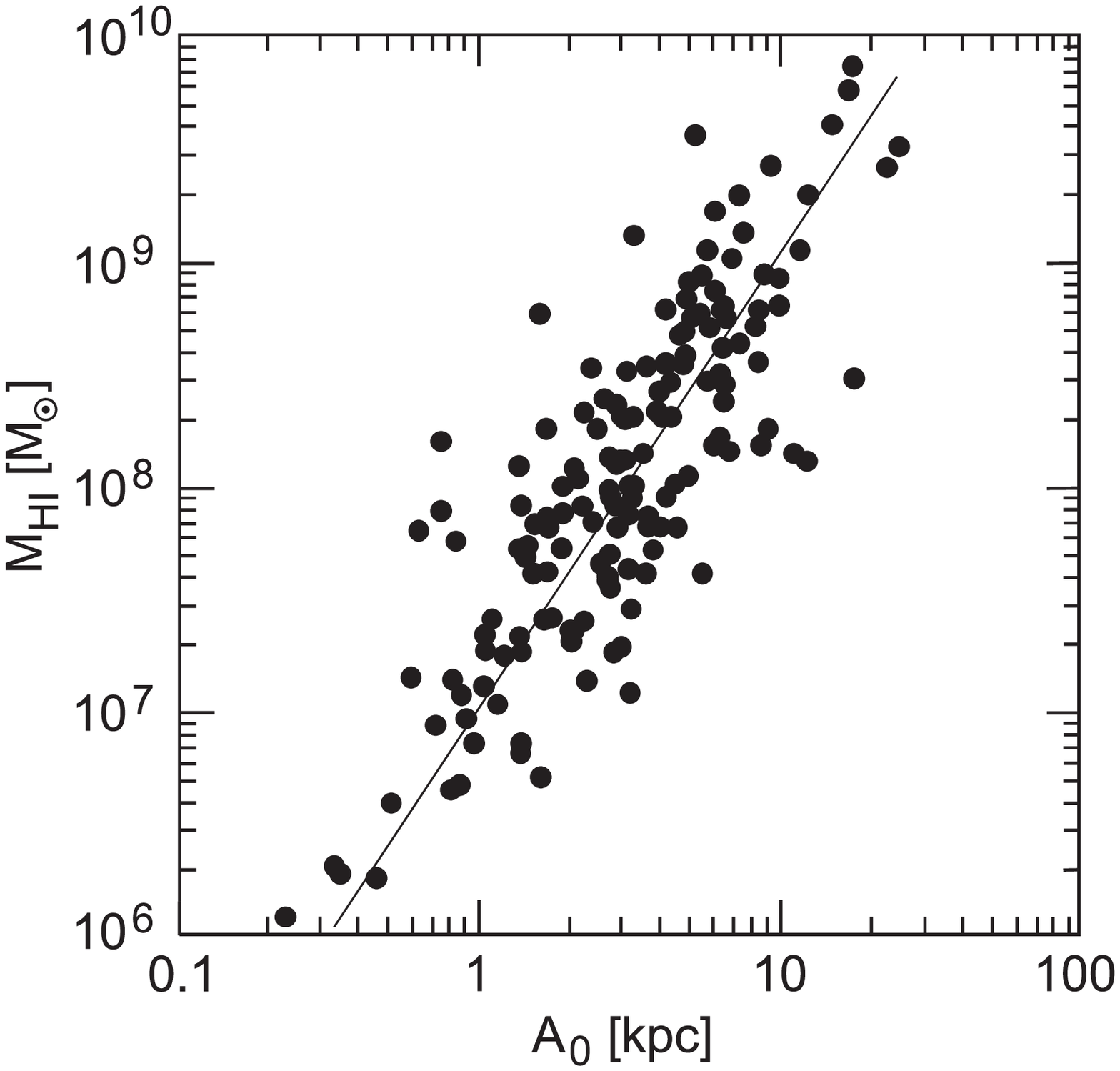,width=8.8cm} 
\caption{ 
     The total mass of neutral hydrogen $M_{HI}$ of the galaxies in our
     sample is plotted versus the linear extent (in kpc). The full line
     represents the regression line for the KKT sample (Huchtmeier \&
     Richter 1988)
       }
\end{figure}

Next we will look at the optical linear diameter $A_{0}$ (in kpc). The 
histogram in Fig.\,5 presents the number of galaxies binned in intervals of 
0.5 kpc width. The
distribution of the optical linear diameters of our galaxies extends from 0.2
kpc to 26 kpc, yet the great majority is smaller than 8 kpc in diameter
(in the de Vaucouleurs $D_{25}$ system). Galaxies in the Local Volume
(indicated by shaded areas) are even smaller with a median value of 
$1.4\pm0.2$ kpc.

Now we will use the correlation of two global parameters to compare
the present sample of galaxies with the previously known galaxies 
in the Local Volume. In Fig.\,6 the total mass of neutral hydrogen $M_{HI}$ of
the galaxies is plotted versus their linear extent $A_{0}$ for this
sample of galaxies.  
The full line is the regression line for the 
KKT sample (Huchtmeier \& Richter 1988). This regression line seems to
be an excellent fit for the present sample, too.
The average HI mass of the galaxies in the Local Volume is 4.6 10$^{7}$
M$_{\sun}$.
\newpage  
The HI masses in Fig.\,6 cover a range from 10$^{6}$ to 10$^{10}$ 
solar masses.  
The HI luminosity function for galaxies  has been studied with
galaxies of 10$^{7}$ and more solar masses in HI so far. 
With the data of the new dwarf galaxies within the Local Volume we will
be able in the end to discuss the HI luminosity function starting 
from 10$^{6}$ solar masses.

\begin{figure}[ht] 
\psfig{figure=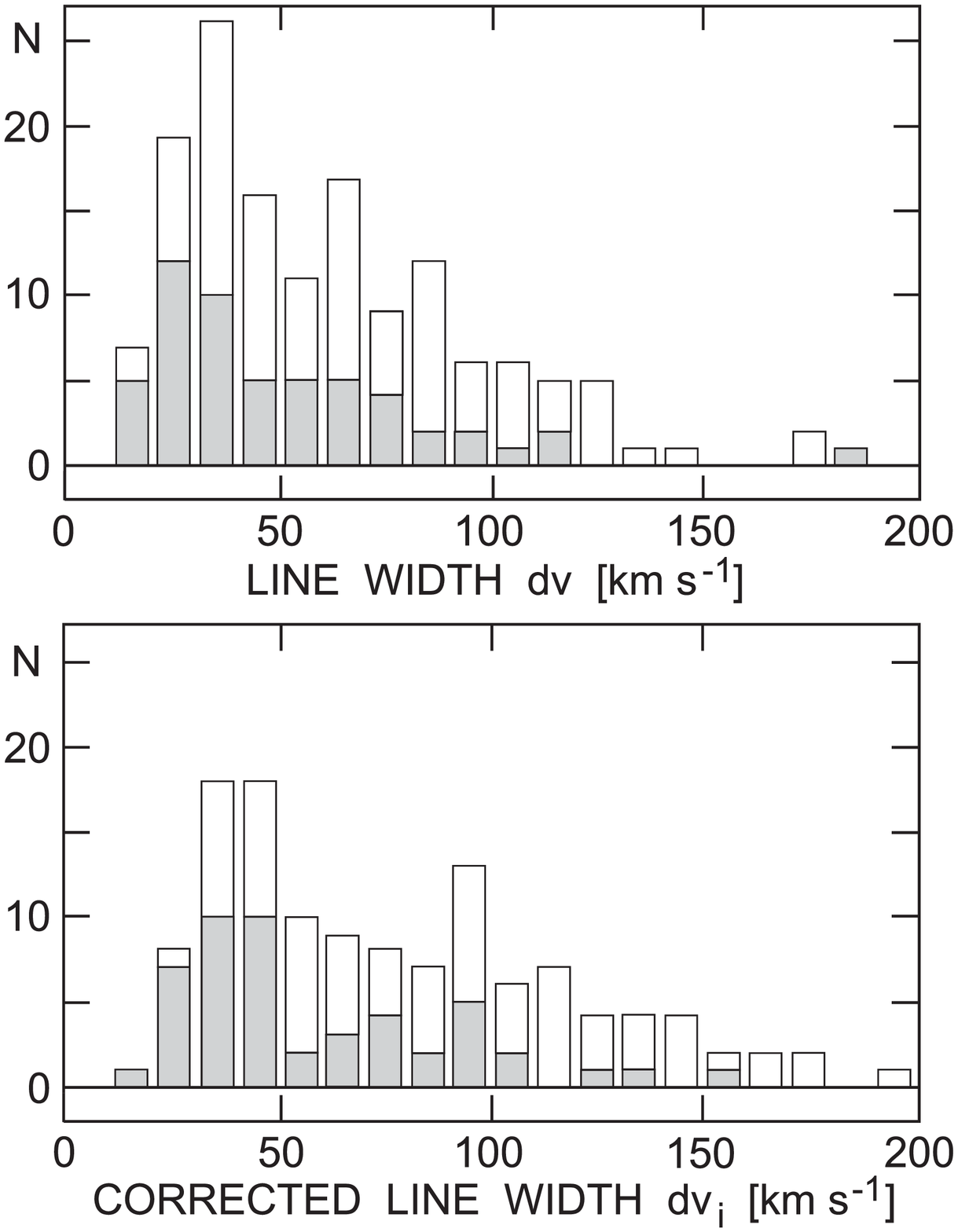,width=8.8cm} 
\caption{
     The distribution of line widths of our galaxy sample is given for
     the observed values ($dv$) in the upper panel and for the (for
     inclination corrected values ($dv_{i}$) in the lower panel. 
     Galaxies within the Local Volume (i.e. within 10 Mpc) are marked 
     by the shaded areas
       }
\end{figure} 

The galaxies in our sample have small line widths on the average.
In Fig.\,7 we present the distribution of observed line widths in the 
upper panel  and the (for inclination) corrected line widths in the
lower panel. The optical axial ratio has been used here to derive the
inclination. Galaxies within the Local Volume are indicated by the
shaded areas. The peak of the line width distribution of the galaxies
within the Local Volume is 39 km\,s$^{-1}$ for the uncorrected and
47 km\,s$^{-1}$ for the corrected line widths. 

The three global parameters we have considered so far point
altogether toward the dwarfish character of the Local Volume objects in our
sample: the average linear diameter of $1.4 \pm 0.2$~kpc (Fig.5), 
the mean total HI mass of 4.6 10$^{7}$ M$_{\sun}$ and the small line
width of less than 50 km\,s$^{-1}$.

\begin{figure}[ht] 
\psfig{figure=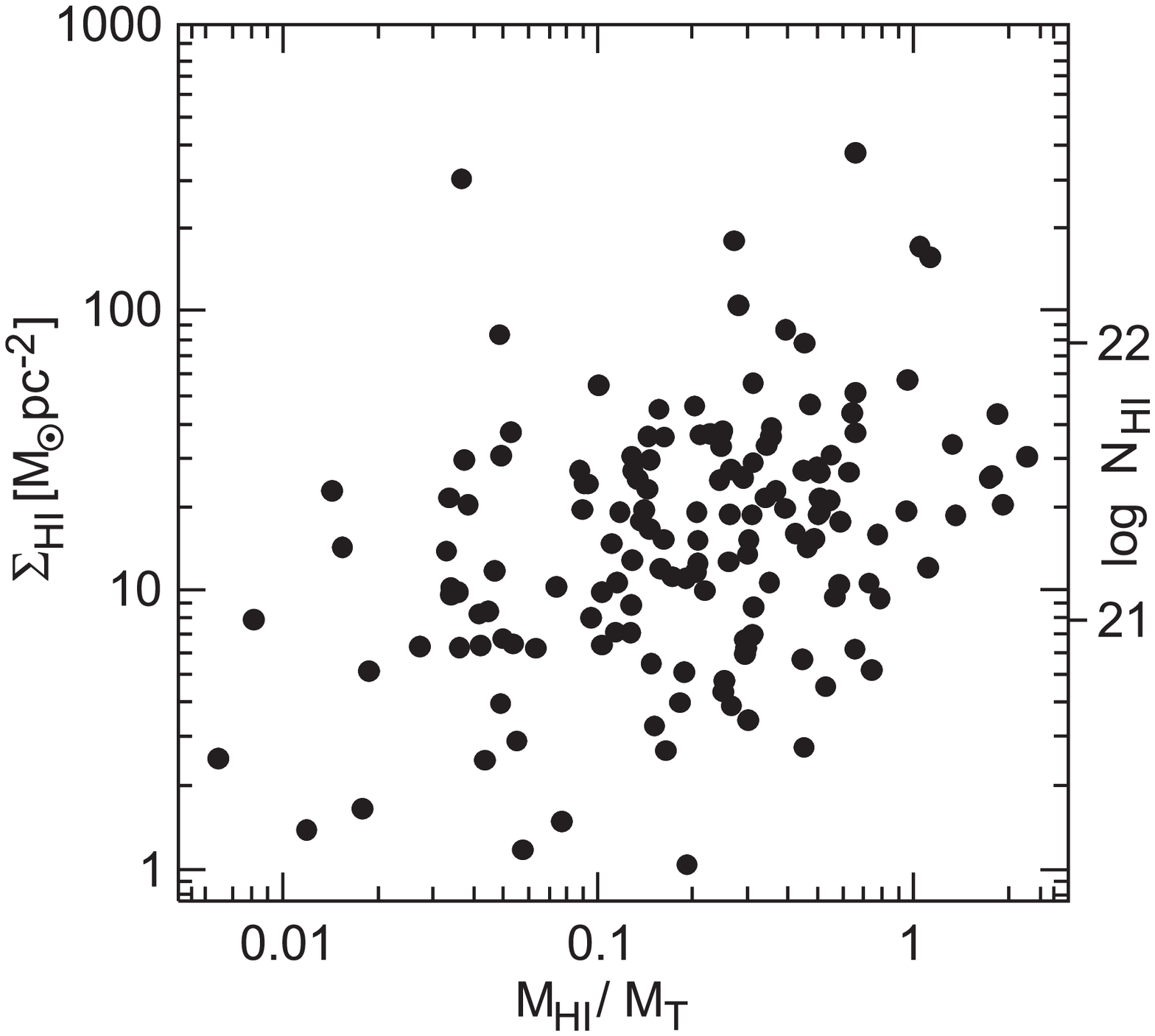,width=8.8cm} 
\caption{
     The pseudo column density of neutral hydrogen ($\Sigma_{HI}$ in
     M$_{\sun}$\,pc$^{-2}$) of our sample as plotted versus the
     relative HI-content ($M_{HI}/M_{T}$). 
       }
\end{figure} 

Two more global parameters are shown in Fig.\,8, pseudo HI surface
density $\Sigma_{HI}$ and the relative HI content $M_{HI}/M_{T}$.
The pseudo HI surface density is obtained by dividing the total HI mass
$M_{HI}$ of the galaxy by the disk area of the galaxy as defined by its
optical diameter $A_{0}$. This quantity is given in units of solar mass
per square parsec as well as in the usual HI column density $N_{HI}$ in 
atoms\, cm$^{-2}$. This quantity is plotted versus the relative HI content
$M_{HI}/M_{T}$. Our galaxies fill the usual range in HI surface density 
as well as in relative HI content as observed for normal galaxies (e.g.
HR).  
The present sample of galaxies is relatively rich in HI. Some of the
scatter in the diagram is due to uncertainties in observed quantities,
especially the inclination which is used to correct the line width 
which itself enters the total mass calculation by the square. The
optical diameters are uncertain for galaxies at low galactic latitudes  
due to the high foreground extinction, e.g. Cas\,2, ESO 137-G27, BK12, ESO 558-11.
If we exclude the confused galaxies and those with heavy galactic
extinction all entries in Fig.\,8 with $\Sigma_{HI} \ge 100$
M$_{\sun}$\,pc$^{-2}$ are gone. Low values of the HI surface density are
not only due to the uncertainties of observational data, the gas content
of dwarf galaxies is very sensitive to outside influences (tidal interactions) 
due to their shallow gravitational potential.

Finally we plot the HI surface brightness versus the optical
surface brightness (Fig.\,9). The surface brightness class (Table 1,
column 7) has been coded from 4 to 1 from high SB to extremely low SB 
in steps of 1. The different errors of the mean values of each class 
essentially depend on the different population size of each SB class.
However, there is a definite trend of the HI surface density to grow
with increasing optical SB by a factor of 2 to 4 (e.g. van der Hulst et 
al. 1993, de Blok 1997).

\begin{figure}[ht] 
\psfig{figure=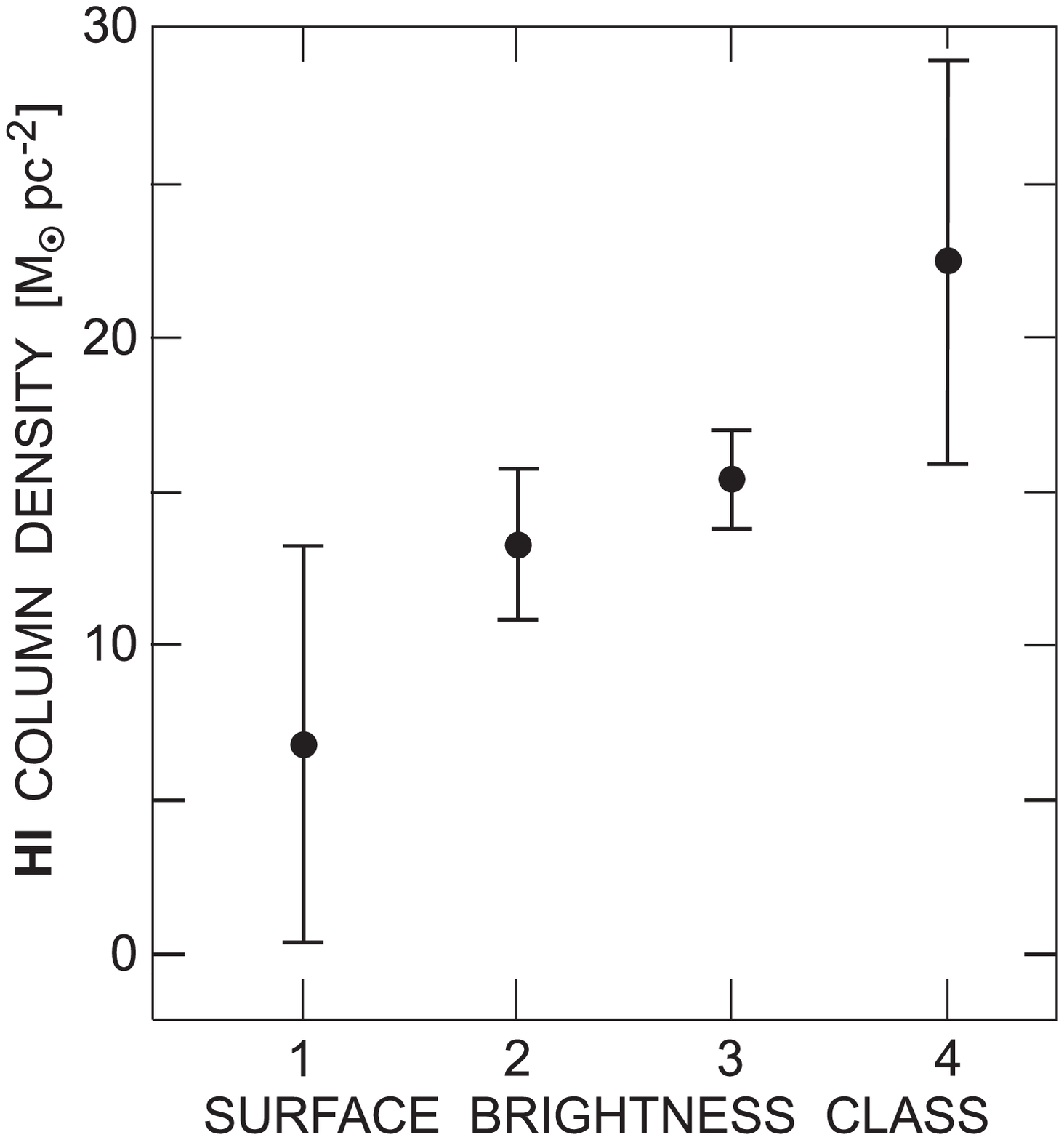,width=8.8cm} 
\caption{
      This figure presents a correlation between the pseudo HI column
      density with the optical surface brightness of the galaxy in our
      actual sample. The surface brightness class is taken from KK98;
      1 = extremely low, 2 = very low, 3 = low, 1 = high SB. The error
      bars correspond to twice the r.m.s. error of the mean of each SB
      class
       }
\end{figure} 

\section{Conclusion}
In this paper we presented an HI search for 257 candidates for nearby
dwarf galaxies. A detection rate of 60\% on the average is quite high 
keeping in mind the 
limited velocity band and the fact that single-dish telescopes are 
literally 'blind' for weak emission in the velocity range of the local
HI emission (i.e. within -140 to +20 km\,s$^{-1}$) and for 
20\%  of HI-poor (spheroidal and Sph/Ir) objects in the sample.
Most of the detected galaxies  are located within the local
supercluster, and about 25\% are members of the Local Volume.
The dwarfs within the Local Volume have a mean linear diameter of 
$1.4\pm0.2$ kpc, a mean observed linewidths of 39 km\,s$^{-1}$, and a mean
total HI mass of 4.6 10$^{7}$ M$_{\odot}$.  
The smallest galaxies have HI masses of just over 10$^{6}$ solar masses.
Once this full-sky survey will be finished we will be able to discuss
the luminosity function of the Local Volume including these tiny dwarf
galaxies. This investigation is especially needed as recent determinations 
of the galaxy luminosity function exhibit an increase for low mass objects. 
The exact value of this increase will be important for deriving the mass 
density in the local universe.  
 
\acknowledgements{
The Australia Telescope is funded by the Commonwealth of Australia for 
operation as a National Facility managed by CSIRO. 

The Nan\c{c}ay Radio Astronomy Observatory is the Unit\'{e} Scientifique
de Nan\c{c}ay  of the Observatoire de Paris, associated as Unit\'{e} de
Service et de Recherche (USR) No. B704 to the French Centre National de
la Recherche Scientifique (CNRS). The Observatory also gratefully
acknowledges the financial support of the Conseil R\'{e}gional of the
R\'{e}gion Centre in France. 

This research has made use of the NASA/IPAC Extragalactic Database
(NED) which is operated by the Jet Propulsion Laboratory, California
Institute of Technology, under contract with the National Aeronautics
and Space Administration. 

This work has been partially supported by the Deutsche
Forschungsgemeinschaft (DFG) under project no. 436 RUS 113/470/0 
and Eh 154/1-1.}

{}
 
\newpage 
\tiny  
\clearpage  
\begin{table*} 
\caption{List of new Local Volume dwarf candidates}
\begin{tabular}{|rcclllllrcl|} \hline
\multicolumn{1}{|r}{KK}&  
\multicolumn{2}{c}{R.A. (1950.0) Dec.}&
\multicolumn{1}{c}{$a$}&
\multicolumn{1}{c}{$b$}&
\multicolumn{1}{l}{Type}&
\multicolumn{1}{c}{S.B.}&
\multicolumn{1}{c}{B$_{t}$}& 
\multicolumn{1}{c}{Ref.(B)}&   
\multicolumn{1}{c}{A$_{b}$}& 
\multicolumn{1}{c|}{Identification} 
 \\ \hline   
\multicolumn{1}{|c}{ }&
\multicolumn{1}{c}{ h m s }&
\multicolumn{1}{c}{\ \ $\degr\ \ \ $\arcmin$ \ \ \ \arcsec$ }&
\multicolumn{1}{c}{arc }&
\multicolumn{1}{c}{min }&
\multicolumn{1}{c}{ }&
\multicolumn{1}{c}{ }&
\multicolumn{1}{c}{ }& 
\multicolumn{1}{c}{ }& 
\multicolumn{1}{c}{ }& 
\multicolumn{1}{c|}{  }  
 \\ \hline
\multicolumn{1}{|c}{1}&
\multicolumn{1}{c}{2}&
\multicolumn{1}{c}{3}&
\multicolumn{1}{c}{4}&
\multicolumn{1}{c}{5}&
\multicolumn{1}{c}{6}&
\multicolumn{1}{c}{7}&
\multicolumn{1}{c}{8}& 
\multicolumn{1}{c}{9}& 
\multicolumn{1}{c}{10}& 
\multicolumn{1}{c|}{11}  
 \\ \hline
&&&&&&& & & & \\
 1 & 00 12 31.6& $-$38 45 43& 0.9& 0.35& Im & H &  & & & AM 0012-384 \\ 
 2 & 00 12 53.5& $-$21 43 17& 2.3& 1.3 & S0 & H & 13.12& NED& 0.06& NGC 59
 \\  
 3 & 00 13 00.0& $-$32 27 36& 1.5& 1.3 & Im & H &  & & & FG 11  \\  
 4& 00 29 27.4& $-$33 32 30& 0.9& 0.8 & Sph/Ir& L & & & & FG 16,AM 0029-333 \\
\hline 
\end{tabular}
\end{table*}
\newpage 
\begin{table*}
\begin{tabular}{|cclllllrccl|} \hline
\multicolumn{1}{|r}{KK}&  
\multicolumn{2}{c}{R.A. (1950.0) Dec.}&
\multicolumn{1}{c}{$a$}&
\multicolumn{1}{c}{$b$}&
\multicolumn{1}{l}{Type}&
\multicolumn{1}{c}{S.B.}&
\multicolumn{1}{c}{B$_{t}$}& 
\multicolumn{1}{c}{Ref.(B)}&   
\multicolumn{1}{c}{A$_{b}$}& 
\multicolumn{1}{c|}{Identification} 
 \\ \hline   
\multicolumn{1}{|c}{ }&
\multicolumn{1}{c}{ h m s }&
\multicolumn{1}{c}{\ \ $\degr\ \ \ \arcmin \ \ \ \arcsec$ }&
\multicolumn{1}{c}{arc }&
\multicolumn{1}{c}{min }&
\multicolumn{1}{c}{ }&
\multicolumn{1}{c}{ }&
\multicolumn{1}{c}{ }& 
\multicolumn{1}{c}{ }& 
\multicolumn{1}{c}{ }& 
\multicolumn{1}{c|}{  }  
 \\ \hline
\multicolumn{1}{|c}{1}&
\multicolumn{1}{c}{2}&
\multicolumn{1}{c}{3}&
\multicolumn{1}{c}{4}&
\multicolumn{1}{c}{5}&
\multicolumn{1}{c}{6}&
\multicolumn{1}{c}{7}&
\multicolumn{1}{c}{8}& 
\multicolumn{1}{c}{9}& 
\multicolumn{1}{c}{10}& 
\multicolumn{1}{c|}{11}  
 \\ \hline
&&&&&&& & & & \\
  269& 03 59 34.8& +71 25 44& 1.2& 0.5& Ir& VL& 17.0& IK& 1.17& BK 21 \\ 
 39& 04 00 06.0& +71 20 00& 1.1:& 0.9:& Ir?& EL& 18.5& IK& 1.12&    \\ 
 40& 04 05 56.0& $-$55 27 21& 1.6& 1.0& Im& H& 14.73& NED& 0.0& AM 0405-552\\
 270& 04 06 44.0& +70 38 33& 0.9& 0.4& Ir& L& 16.6& IK& 1.17&   \\
\hline  
\end{tabular}
\end{table*}
\newpage 
\begin{table*}
\begin{tabular}{|rcclllllrcl|} \hline
\multicolumn{1}{|r}{KK}&  
\multicolumn{2}{c}{R.A. (1950.0) Dec.}&
\multicolumn{1}{c}{$a$}&
\multicolumn{1}{c}{$b$}&
\multicolumn{1}{l}{Type}&
\multicolumn{1}{c}{S.B.}&
\multicolumn{1}{c}{B$_{t}$}& 
\multicolumn{1}{c}{Ref.(B)}&   
\multicolumn{1}{c}{A$_{b}$}& 
\multicolumn{1}{c|}{Identification} 
 \\ \hline   
\multicolumn{1}{|c}{ }&
\multicolumn{1}{c}{ h m s }&
\multicolumn{1}{c}{\ \ $\degr\ \ \ \arcmin \ \ \ \arcsec$ }&
\multicolumn{1}{c}{arc }&
\multicolumn{1}{c}{min }&
\multicolumn{1}{c}{ }&
\multicolumn{1}{c}{ }&
\multicolumn{1}{c}{ }& 
\multicolumn{1}{c}{ }& 
\multicolumn{1}{c}{ }& 
\multicolumn{1}{c|}{  }  
 \\ \hline
\multicolumn{1}{|c}{1}&
\multicolumn{1}{c}{2}&
\multicolumn{1}{c}{3}&
\multicolumn{1}{c}{4}&
\multicolumn{1}{c}{5}&
\multicolumn{1}{c}{6}&
\multicolumn{1}{c}{7}&
\multicolumn{1}{c}{8}& 
\multicolumn{1}{c}{9}& 
\multicolumn{1}{c}{10}& 
\multicolumn{1}{c|}{11}  
 \\ \hline
&&&&&&& & & & \\
81& 09 53 00.8& +68 49 47& 2.6& 2.6& Sph& VL& & & & K 61  \\ 
 82& 10 00 25.3& $-$05 57 55& 0.6& 0.5& Ir& L& 17.73& NED& 0.05&   \\ 
 82& 10 00 25.3& $-$05 57 55& 0.6& 0.5& Ir& L& 17.73& NED& 0.05&   \\ 
 83& 10 01 18.0& +66 48 00& 1.7& 1.7& Sph& VL& & & &  DDO 71,K 63 \\
\hline 
\end{tabular}
\end{table*}
\newpage 
\begin{table*}
\begin{tabular}{|rcclllllrcl|} \hline
\multicolumn{1}{|r}{KK}&  
\multicolumn{2}{c}{R.A. (1950.0) Dec.}&
\multicolumn{1}{c}{$a$}&
\multicolumn{1}{c}{$b$}&
\multicolumn{1}{l}{Type}&
\multicolumn{1}{c}{S.B.}&
\multicolumn{1}{c}{B$_{t}$}& 
\multicolumn{1}{c}{Ref.(B)}&   
\multicolumn{1}{c}{A$_{b}$}& 
\multicolumn{1}{c|}{Identification} 
 \\ \hline   
\multicolumn{1}{|c}{ }&
\multicolumn{1}{c}{ h m s }&
\multicolumn{1}{c}{\ \ $\degr\ \ \ \arcmin \ \ \ \arcsec$ }&
\multicolumn{1}{c}{arc }&
\multicolumn{1}{c}{min }&
\multicolumn{1}{c}{ }&
\multicolumn{1}{c}{ }&
\multicolumn{1}{c}{ }& 
\multicolumn{1}{c}{ }& 
\multicolumn{1}{c}{ }& 
\multicolumn{1}{c|}{  }  
 \\ \hline
\multicolumn{1}{|c}{1}&
\multicolumn{1}{c}{2}&
\multicolumn{1}{c}{3}&
\multicolumn{1}{c}{4}&
\multicolumn{1}{c}{5}&
\multicolumn{1}{c}{6}&
\multicolumn{1}{c}{7}&
\multicolumn{1}{c}{8}& 
\multicolumn{1}{c}{9}& 
\multicolumn{1}{c}{10}& 
\multicolumn{1}{c|}{11}  
 \\ \hline
&&&&&&& & & & \\
127& 12 10 51.5& +30 12 00& 0.9& 0.25& Ir& H& 16.22& UH& 0.04&    \\ 
 128& 12 11 17.1& +05 37 54& 0.6& 0.4& Ir& L&  17.0& IK& 0.0& GR 5    \\
 129& 12 11 38.9& +16 14 35& 1.1& 1.1& Sph& VL& & & & K 88,VCC 108 \\ 
 U7298& 12 14 02.2& +52 30 29& 1.1& 0.6& Ir& L& 15.95& 6m& 0.04&  UGC 7298  \\
\hline  
\end{tabular}
\end{table*}
\newpage 
\begin{table*}
\begin{tabular}{|rcclllllrcl|} \hline
\multicolumn{1}{|r}{KK}&  
\multicolumn{2}{c}{R.A. (1950.0) Dec.}&
\multicolumn{1}{c}{$a$}&
\multicolumn{1}{c}{$b$}&
\multicolumn{1}{l}{Type}&
\multicolumn{1}{c}{S.B.}&
\multicolumn{1}{c}{B$_{t}$}& 
\multicolumn{1}{c}{Ref.(B)}&   
\multicolumn{1}{c}{A$_{b}$}& 
\multicolumn{1}{c|}{Identification} 
 \\ \hline   
\multicolumn{1}{|c}{ }&
\multicolumn{1}{c}{ h m s }&
\multicolumn{1}{c}{\ \ $\degr\ \ \ \arcmin \ \ \ \arcsec$ }&
\multicolumn{1}{c}{arc }&
\multicolumn{1}{c}{min }&
\multicolumn{1}{c}{ }&
\multicolumn{1}{c}{ }&
\multicolumn{1}{c}{ }& 
\multicolumn{1}{c}{ }& 
\multicolumn{1}{c}{ }& 
\multicolumn{1}{c|}{  }  
 \\ \hline
\multicolumn{1}{|c}{1}&
\multicolumn{1}{c}{2}&
\multicolumn{1}{c}{3}&
\multicolumn{1}{c}{4}&
\multicolumn{1}{c}{5}&
\multicolumn{1}{c}{6}&
\multicolumn{1}{c}{7}&
\multicolumn{1}{c}{8}& 
\multicolumn{1}{c}{9}& 
\multicolumn{1}{c}{10}& 
\multicolumn{1}{c|}{11}  
 \\ \hline
&&&&&&& & & & \\
 167& 12 49 17.8& +26 22 56& 0.8& 0.5& Ir?& L& 16.7& IK& 0.02&   \\ 
 168& 12 50 37.8& +03 42 44& 0.5& 0.4& Ir?& VL& 17.3& IK& 0.0&    \\ 
 169& 12 50 41.3& +12 54 24& 0.8& 0.6& Ir& L& 17.1& IK& 0.05&   \\ 
 170& 12 52 11.5& $-$28 04 12& 0.8& 0.6& Im& L& 17.01& NED& 0.30& AM 1252-280 \\
\hline 
\end{tabular}
\end{table*}
\newpage 
\begin{table*} 
\begin{tabular}{|rcclllllrcl|} \hline
\multicolumn{1}{|r}{KK}&  
\multicolumn{2}{c}{R.A. (1950.0) Dec.}&
\multicolumn{1}{c}{$a$}&
\multicolumn{1}{c}{$b$}&
\multicolumn{1}{l}{Type}&
\multicolumn{1}{c}{S.B.}&
\multicolumn{1}{c}{B$_{t}$}& 
\multicolumn{1}{c}{Ref.(B)}&   
\multicolumn{1}{c}{A$_{b}$}& 
\multicolumn{1}{c|}{Identification} 
 \\ \hline   
\multicolumn{1}{|c}{ }&
\multicolumn{1}{c}{ h m s }&
\multicolumn{1}{c}{\ \ $\degr\ \ \ \arcmin \ \ \ \arcsec$ }&
\multicolumn{1}{c}{arc }&
\multicolumn{1}{c}{min }&
\multicolumn{1}{c}{ }&
\multicolumn{1}{c}{ }&
\multicolumn{1}{c}{ }& 
\multicolumn{1}{c}{ }& 
\multicolumn{1}{c}{ }& 
\multicolumn{1}{c|}{  }  
 \\ \hline
\multicolumn{1}{|c}{1}&
\multicolumn{1}{c}{2}&
\multicolumn{1}{c}{3}&
\multicolumn{1}{c}{4}&
\multicolumn{1}{c}{5}&
\multicolumn{1}{c}{6}&
\multicolumn{1}{c}{7}&
\multicolumn{1}{c}{8}& 
\multicolumn{1}{c}{9}& 
\multicolumn{1}{c}{10}& 
\multicolumn{1}{c|}{11}  
 \\ \hline
&&&&&&& & & & \\
 220& 13 45 22.2& +33 27 25& 0.8:& 0.7:& Ir& L& 16.5& IK& 0.0&    \\ 
 223& 13 46 07.8& +40 48 08& 0.4& 0.3& Ir?& L& 17.4& IK& 0.0&   \\ 
 224& 13 46 52.8& +43 50 54& 1.0& 0.4& Ir& L& 16.9& IK& 0.0&    \\ 
 225& 13 52 51.5& +37 55 42& 0.45& 0.3& Ir& L& & & &   \\
\hline
\end{tabular}
\end{table*}
\newpage 
\clearpage  
\tiny  
\begin{table*}
\caption{List of new Local Volume dwarf candidates}
\begin{tabular}{|rccllclccl|} \hline
\multicolumn{1}{|r}{KK}&  
\multicolumn{1}{c}{HI-flux}&
\multicolumn{1}{c}{S$_{max}$}&
\multicolumn{1}{c}{velocity}&
\multicolumn{1}{l}{line width}&
\multicolumn{1}{c}{distance}&
\multicolumn{1}{c}{M$_{Bt}$}& 
\multicolumn{1}{c}{HI mass}&   
\multicolumn{1}{c}{M$_{HI}$/L$_{B}$}& 
\multicolumn{1}{c|}{Comments}
\\ \hline   
\multicolumn{1}{|c}{No. }&
\multicolumn{1}{c}{Jy$\,$km$\,$s$^{-1}$}&
\multicolumn{1}{c}{mJy }&
\multicolumn{1}{c}{km$\,$s$^{-1}$}&
\multicolumn{1}{c}{km$\,$s$^{-1}$}&
\multicolumn{1}{c}{Mpc }&
\multicolumn{1}{c}{ }& 
\multicolumn{1}{c}{10$^{7}$M$_{\odot}$}&
\multicolumn{1}{c}{ }& 
\multicolumn{1}{c|}{  }  
 \\ \hline 
\multicolumn{1}{|c}{1}&
\multicolumn{1}{c}{2}&
\multicolumn{1}{c}{3}&
\multicolumn{1}{c}{4}&
\multicolumn{1}{c}{5}&
\multicolumn{1}{c}{6}&
\multicolumn{1}{c}{7}&
\multicolumn{1}{c}{8}&
\multicolumn{1}{c}{9}&
\multicolumn{1}{c|}{10} 
 \\ \hline  
&&&&&& & & & \\
 1 &  & $\pm$15 & & & & & & &ATCA \\ 
2 & 2.66& 57$\pm$5.9& 361$\pm$2& 50 62 76& 5.4& -15.70& 1.9& 0.06& \\  
 3 & &  $\pm24$& & & & & & &N \\  
 4& &   $\pm$35& & & & & & &N  \\
\hline
\end{tabular}
\end{table*}
\newpage 
\begin{table*}
\begin{tabular}{|rccllclccl|} \hline
\multicolumn{1}{|r}{KK}&  
\multicolumn{1}{c}{HI-flux}&
\multicolumn{1}{c}{S$_{max}$}&
\multicolumn{1}{c}{velocity}&
\multicolumn{1}{l}{line width}&
\multicolumn{1}{c}{distance}&
\multicolumn{1}{c}{M$_{Bt}$}& 
\multicolumn{1}{c}{HI mass}&   
\multicolumn{1}{c}{M$_{HI}$/L$_{B}$}& 
\multicolumn{1}{c|}{Comment}
\\ \hline   
\multicolumn{1}{|c}{No. }&
\multicolumn{1}{c}{Jy$\,$km$\,$s$^{-1}$}&
\multicolumn{1}{c}{mJy }&
\multicolumn{1}{c}{km$\,$s$^{-1}$}&
\multicolumn{1}{c}{km$\,$s$^{-1}$}&
\multicolumn{1}{c}{Mpc }&
\multicolumn{1}{c}{ }& 
\multicolumn{1}{c}{10$^{7}$M$_{\odot}$}&
\multicolumn{1}{c}{  }&  
\multicolumn{1}{c|}{  }  
\\ \hline
\multicolumn{1}{|c}{1}&
\multicolumn{1}{c}{2}&
\multicolumn{1}{c}{3}&
\multicolumn{1}{c}{4}&
\multicolumn{1}{c}{5}&
\multicolumn{1}{c}{6}&
\multicolumn{1}{c}{7}&
\multicolumn{1}{c}{8}&
\multicolumn{1}{c}{9}&
\multicolumn{1}{c|}{10} 
 \\ \hline  
 &&&&&& & & & \\
 269& 6.83& 69$\pm$5& 1734$\pm$2& 120 133 136& 25.8& -16.48& 110& 1.75&  \\  
 39& 2.0& 21$\pm$3.5& 1732$\pm$4& 123 125 126& 25.7& -14.73& 17& 1.42&  \\  
 40& 2.87&53$\pm$5& 1066$\pm$2& 101 109 112 & 11.4& -15.69& 8.8& 0.30&ATCA \\  
 270& 4.6& 110$\pm$5& 1159$\pm$1& 42 57 59& 18.1& -16.09& 36& 0.85&  \\
\hline
\end{tabular}
\end{table*}
\newpage  
\begin{table*}
\begin{tabular}{|rccllclccl|} \hline
\multicolumn{1}{|r}{KK}&  
\multicolumn{1}{c}{HI-flux}&
\multicolumn{1}{c}{S$_{max}$}&
\multicolumn{1}{c}{velocity}&
\multicolumn{1}{l}{line width}&
\multicolumn{1}{c}{distance}&
\multicolumn{1}{c}{M$_{Bt}$}& 
\multicolumn{1}{c}{HI mass}&   
\multicolumn{1}{c}{M$_{HI}$/L$_{B}$}& 
\multicolumn{1}{c|}{Comment}
\\ \hline   
\multicolumn{1}{|c}{No. }&
\multicolumn{1}{c}{Jy$\,$km$\,$s$^{-1}$}&
\multicolumn{1}{c}{mJy }&
\multicolumn{1}{c}{km$\,$s$^{-1}$}&
\multicolumn{1}{c}{km$\,$s$^{-1}$}&
\multicolumn{1}{c}{Mpc }&
\multicolumn{1}{c}{ }& 
\multicolumn{1}{c}{10$^{8}$M$_{\odot}$}&
\multicolumn{1}{c}{10$^{9}$M$_{\odot}$}& 
\multicolumn{1}{c|}{  }  
 \\ \hline
\multicolumn{1}{|c}{1}&
\multicolumn{1}{c}{2}&
\multicolumn{1}{c}{3}&
\multicolumn{1}{c}{4}&
\multicolumn{1}{c}{5}&
\multicolumn{1}{c}{6}&
\multicolumn{1}{c}{7}&
\multicolumn{1}{c}{8}&
\multicolumn{1}{c}{9}&
\multicolumn{1}{c|}{10} 
 \\ \hline  
&&&&&& & & & \\
 81& &$\pm8.5$ & & & & & & &                      \\
 82& 3.94& 165$\pm$6& 185$\pm$1& 21 31 35& & & & &  \\ 
 82& 1.54& 40$\pm$3.2& 3769$\pm$6& 35 48 55&  47.3& -15.75& 84& 2.53&    \\    
 83& &$\pm$6.2& & & & & & &                   \\
\hline
\end{tabular}
\end{table*}
\newpage 
\begin{table*}
\begin{tabular}{|rccllclccl|} \hline
\multicolumn{1}{|r}{KK}&  
\multicolumn{1}{c}{HI-flux}&
\multicolumn{1}{c}{S$_{max}$}&
\multicolumn{1}{c}{velocity}&
\multicolumn{1}{l}{line width}&
\multicolumn{1}{c}{distance}&
\multicolumn{1}{c}{M$_{Bt}$}& 
\multicolumn{1}{c}{HI mass}&   
\multicolumn{1}{c}{M$_{HI}$/L$_{B}$}&  
\multicolumn{1}{r|}{Comments} 
\\ \hline   
\multicolumn{1}{|c}{No. }&
\multicolumn{1}{c}{Jy$\,$km$\,$s$^{-1}$}&
\multicolumn{1}{c}{mJy }&
\multicolumn{1}{c}{km$\,$s$^{-1}$}&
\multicolumn{1}{c}{km$\,$s$^{-1}$}&
\multicolumn{1}{c}{Mpc }&
\multicolumn{1}{c}{ }& 
\multicolumn{1}{c}{10$^{7}$M$_{\odot}$}&
\multicolumn{1}{c}{}& 
\multicolumn{1}{c|}{  }  
 \\ \hline
\multicolumn{1}{|c}{1}&
\multicolumn{1}{c}{2}&
\multicolumn{1}{c}{3}&
\multicolumn{1}{c}{4}&
\multicolumn{1}{c}{5}&
\multicolumn{1}{c}{6}&
\multicolumn{1}{c}{7}&
\multicolumn{1}{c}{8}&
\multicolumn{1}{c}{9}&
\multicolumn{1}{c|}{10} 
 \\ \hline  
&&&&&& & & & \\
 127& 2.87& 41$\pm$4.4& 131$\pm$6& 64 112 117& 1.6& -10.23& 0.18& 0.92& \\ 
 128& 1.80& 24$\pm$3.2& 1690$\pm$2& 56 73 76& 21.0& -14.72& 19& 1.58& \\
 129& & $\pm$6.1& & & & & & &  \\ 
 U7298& 5.44& 193$\pm$6.5& 172$\pm$1& 26 38 41& 3.5& -12.01& 1.6& 1.66& \\
\hline
\end{tabular}
\end{table*}
\newpage 
\begin{table*}
\begin{tabular}{|rccllclccl|} \hline
\multicolumn{1}{|r}{KK}&  
\multicolumn{1}{c}{HI-flux}&
\multicolumn{1}{c}{S$_{max}$}&
\multicolumn{1}{c}{velocity}&
\multicolumn{1}{l}{line width}&
\multicolumn{1}{c}{distance}&
\multicolumn{1}{c}{M$_{Bt}$}& 
\multicolumn{1}{c}{HI mass}&   
\multicolumn{1}{c}{M$_{HI}$/L$_{B}$}& 
\multicolumn{1}{c|}{Comments} 
\\ \hline   
\multicolumn{1}{|c}{No. }&
\multicolumn{1}{c}{Jy$\,$km$\,$s$^{-1}$}&
\multicolumn{1}{c}{mJy }&
\multicolumn{1}{c}{km$\,$s$^{-1}$}&
\multicolumn{1}{c}{km$\,$s$^{-1}$}&
\multicolumn{1}{c}{Mpc }&
\multicolumn{1}{c}{ }& 
\multicolumn{1}{c}{10$^{7}$M$_{\odot}$}&
\multicolumn{1}{c}{}& 
\multicolumn{1}{c|}{  }  
 \\ \hline
\multicolumn{1}{|c}{1}&
\multicolumn{1}{c}{2}&
\multicolumn{1}{c}{3}&
\multicolumn{1}{c}{4}&
\multicolumn{1}{c}{5}&
\multicolumn{1}{c}{6}&
\multicolumn{1}{c}{7}&
\multicolumn{1}{c}{8}&
\multicolumn{1}{c}{9}&
\multicolumn{1}{c|}{10} 
 \\ \hline  
&&&&&& & & & \\
 167& 1.99& 51$\pm$3.2& 1247$\pm$4& 37 50 52& 16.6& -14.55& 13& 1.24& \\ 
 168& 1.35& 45$\pm$4.3& 2785$\pm$4& 39 41 54& 35.8& -15.53& 33& 1.31& \\
 169& 1.65& 39$\pm$4.2& 1818$\pm$2& 46 54 55& 23.4& -14.88& 20& 1.40& \\
 170& 1.71&43$\pm$4.3 & 641$\pm$3& 35 55 59& 6.5& -12.57& 1.7& 1.04&  \\
\hline
\end{tabular}
\end{table*}
\newpage 
\begin{table*} 
\begin{tabular}{|rccllclccl|} \hline
\multicolumn{1}{|r}{KK}&  
\multicolumn{1}{c}{HI-flux}&
\multicolumn{1}{c}{S$_{max}$}&
\multicolumn{1}{c}{velocity}&
\multicolumn{1}{l}{line width}&
\multicolumn{1}{c}{distance}&
\multicolumn{1}{c}{M$_{Bt}$}& 
\multicolumn{1}{c}{HI mass}&   
\multicolumn{1}{c}{M$_{HI}$/L$_{B}$}& 
\multicolumn{1}{l|}{Comments} 
\\ \hline   
\multicolumn{1}{|c}{No. }&
\multicolumn{1}{c}{Jy$\,$km$\,$s$^{-1}$}&
\multicolumn{1}{c}{mJy }&
\multicolumn{1}{c}{km$\,$s$^{-1}$}&
\multicolumn{1}{c}{km$\,$s$^{-1}$}&
\multicolumn{1}{c}{Mpc }&
\multicolumn{1}{c}{ }& 
\multicolumn{1}{c}{10$^{7}$M$_{\odot}$}&
\multicolumn{1}{c}{}& 
\multicolumn{1}{c|}{  }  
 \\ \hline
\multicolumn{1}{|c}{1}&
\multicolumn{1}{c}{2}&
\multicolumn{1}{c}{3}&
\multicolumn{1}{c}{4}&
\multicolumn{1}{c}{5}&
\multicolumn{1}{c}{6}&
\multicolumn{1}{c}{7}&
\multicolumn{1}{c}{8}&
\multicolumn{1}{c}{9}&
\multicolumn{1}{c|}{10} 
 \\ \hline  
&&&&&& & & & \\
 220& 0.8& 33$\pm$5.6& 769$\pm$2& 27 33 34& 11.1& -13.76& 2.3& 0.47& \\
 223& 4.26& 53$\pm$5.5& 2594$\pm$8& 69 130 138& 35.8& -15.45& 130& 5.51& \\ 
 224& 0.84& 25$\pm$6.7& 1156$\pm$6& 26 43 54& 16.8& -14.49& 5.3& 0.55& \\ 
 225& & $\pm$6.3& & & & & & &        \\
\hline
\end{tabular}
\end{table*}

\end{document}